\documentclass[reprint,superscriptaddress,amsmath,amssymb,aip,jcp,floatfix]{revtex4-1}
\usepackage[utf8]{inputenc}
\usepackage{graphicx}
\usepackage{enumitem}
\usepackage{textcomp}
\usepackage{gensymb}
\usepackage{amsmath}
\usepackage{amsthm}
\usepackage{amssymb}
\usepackage{dcolumn}
\usepackage{color}

\usepackage[normalem]{ulem}

\usepackage{multirow}
\usepackage{algorithm}
\usepackage{algpseudocode}
\usepackage{parskip}
\usepackage{makecell}

\begin{document}

\title{Interatomic force from neural network based variational quantum Monte Carlo}

\author{Yubing Qian}
\affiliation{School of Physics, Peking University, Beijing 100871, People’s Republic of China}

\author{Weizhong Fu}
\affiliation{School of Physics, Peking University, Beijing 100871, People’s Republic of China}
\affiliation{ByteDance Inc, Zhonghang Plaza, No. 43,  North 3rd Ring West Road, Haidian District, Beijing.}

\author{Weiluo Ren}
\affiliation{ByteDance Inc, Zhonghang Plaza, No. 43,  North 3rd Ring West Road, Haidian District, Beijing.}

\author{Ji Chen}
\email{ji.chen@pku.edu.cn}
\affiliation{School of Physics, Peking University, Beijing 100871, People’s Republic of China}
\affiliation{
    Collaborative Innovation Center of Quantum Matter, Beijing 100871, People’s Republic of China
}
\affiliation{Interdisciplinary Institute of Light-Element Quantum Materials and Research Center for Light-Element Advanced Materials, Peking University, Beijing 100871, People’s Republic of China
}
\affiliation{Frontiers Science Center for Nano-Optoelectronics, Peking University, Beijing 100871, People’s Republic of China}

\date{\today}

\begin{abstract}
    Accurate \textit{ab initio} calculations are of fundamental importance in physics, chemistry, biology, and materials science, which have witnessed rapid development in the last couple of years with the help of machine learning computational techniques such as neural networks.
    Most of the recent efforts applying neural networks to \textit{ab initio} calculation have been focusing on the energy of the system.
    In this study, we take a step forward and look at the interatomic force obtained with neural network wavefunction methods by implementing and testing several commonly used force estimators in variational quantum Monte Carlo (VMC).
    Our results show that neural network ansatz can improve the calculation of interatomic force upon traditional VMC.
    The relation between the force error and the quality of neural network, the contribution of different force terms, and the computational cost of each term are also discussed to provide guidelines for future applications.
    Our work demonstrates it is promising to apply neural network wavefunction methods in simulating structures/dynamics of molecules/materials and provide training data for developing accurate force fields.
\end{abstract}

\maketitle

\section{Introduction}

\textit{Ab initio} methods have been important tools to calculate the electronic structure of molecules and materials over the past few decades, closing the gap between fundamental theories and experimental measurements \cite{martin_2004}.
Typical \textit{ab initio} methods include density functional theory (DFT) \cite{KohnNobel}, post Hartree-Fock wavefunction approaches such as coupled cluster (CC) and configuration interaction (CI) \cite{PopleNobel}, and statistical methods such as quantum Monte Carlo (QMC) \cite{FoulkesReview}.
However, even though many methods are widely referred to as \textit{ab initio} or first principles methods, it's worth mentioning that there are many approximations and empirical methods applied, trading off accuracy for efficiency, since the many-electron Schrödinger equation is nearly impossible to be solved exactly.
Recently, with the development of computational capabilities and efficient algorithms, deep neural networks have shown great potential to push \textit{ab initio} calculations towards the exact \cite{han_solving_2019,pfau_ab_2020,choo_fermionic_2020,PauliNet,kirkpatrick_pushing_2021,deeperwin,li_ab_2022,ren_towards_2022}.
The Fermionic Neural Network (FermiNet) developed by Pfau et al. is one of those promising approaches which can provide accurate ground state energy of molecules \cite{pfau_ab_2020,spencer_better_2020,deepmind_ferminet_2022}.

The use of FermiNet, or similarly other neural networks, is often integrated with the variational quantum Monte Carlo (VMC) approach.
The neural network plays the role of the trial wavefunction ansatz, and VMC is employed to train the neural network and to estimate physical quantities, which in some sense is similar to traditional QMC methods \cite{needs2020variational,kent_qmcpack_2020,nakano_turborvb_2020}.
QMC performs well in providing accurate energy results,
but it requires a lot of extra work to get a good result for other observables, e.g. interatomic forces and dipoles \cite{van_rhijn_energy_2022}.
That is mainly because straightforward estimate results in infinite variance and numerical instabilities in Monte Carlo simulations.
However, from the perspective of materials modelling, it is also necessary to compute those quantities.
For example, forces are needed for relaxing the structure, performing molecular dynamics, and training force fields.
In traditional QMC simulations, different proposals have been put forward in the past few decades\cite{filippi_correlated_2000,assaraf_computing_2000,assaraf_zero-variance_2003,trail_heavy_2008,attaccalite_stable_2008,badinski_methods_2010,moroni_practical_2014,pathak_light_2020}, but these methods have not been applied and tested in neural network based QMC.
In FermiNet for example, the cusp conditions are approximated instead of written exactly, and there is no guarantee that FermiNet can obtain the correct interatomic force using the estimators proposed for traditional ansatz.
It is also worth noting that the performances of these estimators are to some extent related to the variance and the bias in practical quantum Monte Carlo simulations, hence it is not clear whether the more accurate neural network ansatz would alter the performance of different force estimators.


In this work, we focus on the interatomic force from FermiNet-based VMC simulation (FermiNet-VMC).
There are other network based wavefunction ansatz that combines neural networks with traditional ansatz, and we expect the study of force estimators in FermiNet can also provide insights to the interatomic force calculations with other networks \cite{PauliNet,gerard_gold-standard_2022}.
We implement the modified estimators suggested in traditional variation quantum Monte Carlo
and obtain forces of several diatomic molecules along their potential energy curves
\cite{assaraf_zero-variance_2003,umrigar_two_1989,filippi_correlated_2000}.
Our study shows that the improvement to the wavefunction by the neural network ansatz can also benefit the force calculation, which promises the promotion of neural network approaches in future modelling.
We also test different estimators on neural network wavefunction trained at different levels and examine the relationship between the accuracy of force and the quality of neural network wavefunction.

\section{Methods and computational details}

\subsection{FermiNet-VMC method}

Solving the Schrödinger equation of electrons is a fundamental task in electronic structure calculation.
Under the Born-Oppenheimer approximation, the motion of nuclei is ignored and the equation can be written as
\begin{equation}
    H \psi(\mathbf{x}_1,\mathbf{x}_2,\cdots,\mathbf{x}_{n_{\text{elec}}}) = E \psi(\mathbf{x}_1,\mathbf{x}_2,\cdots,\mathbf{x}_{n_{\text{elec}}})
\end{equation}
where the Hamiltonian
\begin{equation}
    \begin{aligned}
        H =& - \frac{1}{2} \sum_{i=1}^{n_{\text{elec}}} \nabla_i^2 + \sum_{i=1}^{n_{\text{elec}}} \sum_{j=1}^{i-1} \frac{1}{|\mathbf{r}_i - \mathbf{r}_j|} \\ 
        &- \sum_{i=1}^{n_{\text{elec}}} \sum_{I=1}^{n_{\text{atom}}} \frac{Z_I}{|\mathbf{r}_i - \mathbf{R}_I|} + \sum_{I=1}^{n_{\text{atom}}} \sum_{J=1}^{I-1} \frac{Z_I Z_J}{|\mathbf{R}_I-\mathbf{R}_J|}
    \end{aligned}
\end{equation}
and $\mathbf{x}_i = (\mathbf{r}_i, \sigma_i)$ is the spatial and spin coordinates of electron $i$.
Electrons obey Fermi-Dirac statistics and the wavefunction should be antisymmetric under the permutation of $(\mathbf{x}_1,\mathbf{x}_2,\cdots,\mathbf{x}_{n_{\text{elec}}})$. 
The constraint makes the expression of many-body wavefunction even harder, thus prohibiting its solution.

In general, if one expresses the wavefunction with 
parameters $\theta$, and outputs an antisymmetric wavefunction, then
it is possible to use the energy expectation value of the ground state $E_{\theta}$ as the loss
function to optimize the wavefunction towards the ground state.
\begin{equation}
    E_{\theta}=\frac{\langle\psi_{\theta}|H|\psi_{\theta}\rangle}{\langle\psi_{\theta}| \psi_{\theta}\rangle}
\end{equation}
The above integral can be calculated using the Monte Carlo method, and such an approach is often regarded as the VMC method.

FermiNet is an antisymmetric neural network, which expresses the many-body wavefunction in replacement for traditional wavefunction ansatz.
Each orbital in the determinant not only depends on the coordinates of a single electron but all the electrons.
The neural network is designed so that the ansatz can be more expressive if more layers and hidden units are added into the neural network. 
To acquire an expressive neural network ansatz, the number of parameters $\theta$ is often chosen of order $10^5$ to $10^6$, much larger than the number in traditional wavefunction ansatz.
With so many parameters, solving the linear system for natural gradient exactly is impossible, so a modified version of Kronecker-factored approximate curvature (KFAC)\cite{pmlr-v37-martens15} is applied for the neural network to minimize $E_{\theta}$. And finally, the ground state wavefunction can be obtained.
High accuracy can be achieved as long as the network is expressive enough and optimizations are sufficiently performed.

\subsection{Force estimators}
\subsubsection{Bare estimator}

In VMC, the variational principle ensures the energy estimation has an error proportional to the square of wavefunction error, namely
\begin{equation}
    \Delta_{E}=\frac{\left\langle\psi_{T}-\psi_{0}\left|H-E_{0}\right| \psi_{T}-\psi_{0}\right\rangle}{\left\langle\psi_{T} \mid \psi_{T}\right\rangle} \sim O[(\psi_T - \psi_0)^2]
\end{equation}
However, if we consider a bare force estimator on the nucleus of the atom $A$
\begin{equation}
\begin{aligned}
    \mathbf{F}_{A}&=\mathbf{F}_{A,\text{aa}}+\mathbf{F}_{A,\text{ae}}\\
    &= Z_{A} \sum_{\substack{j=1 \\ j\ne A}}^{n_\text{atom}} Z_j \frac{\mathbf{R}_{A}-\mathbf{R}_{j}}{|\mathbf{R}_{A}-\mathbf{R}_{j}|^3}
    + Z_A\sum_{i=1}^{n_{\text{elec}}}\frac{\mathbf{r}_{i}-\mathbf{R}_A}{\left|\mathbf{r}_{i}-\mathbf{R}_A\right|^{3}}
\end{aligned}
\end{equation}
and sample it by trial wave function $\psi_T^2$ instead of the exact wave function $\psi_0^2$,
the systematic error behaves like $\langle \mathbf{F}_A \rangle_{\psi_T^2} - \langle \mathbf{F}_A \rangle_{\psi_0^2} \sim O[\psi_T - \psi_0]$ and the variance behaves like $O[1]$\cite{assaraf_zero-variance_2003}.
Therefore, the bare force estimator would lead to a very slow convergence of force in practice with both large systematic and statistical errors.
To solve this problem regularized force estimators for VMC have been proposed in several seminal works.
In this work, we have implemented both the bare estimator and several regularized estimators based on FermiNet-VMC, which are described below.

\subsubsection{Assaraf-Caffarel estimator}

Assaraf and Caffarel suggested a regularized estimator\cite{assaraf_zero-variance_2003}:
\begin{equation}\label{general-O-estimator}
    \widetilde{\mathbf{F}}_{A,\text{AC-ZVZB}} \equiv \mathbf{F}_{A}-\frac{\left(H-E_{L}\right) \widetilde{\boldsymbol{\psi}}_{A}}{\psi_{T}}+2\left(E_{v}-E_{L}\right) \frac{\widetilde{\boldsymbol{\psi}}_{A}}{\psi_{T}}
\end{equation}
where $E_L$ is the local energy
\begin{equation}
    E_L=\frac{H\psi_T}{\psi_T}
\end{equation}
and $E_v$ is the mean energy sampled over probability density $\psi_T^2$
\begin{equation}
    E_v=\left\langle E_L \right\rangle_{\psi_T^2}
\end{equation}
and $\widetilde{\boldsymbol{\psi}}_{A}$ is the trial function for $\dfrac{\partial\psi_0}{\partial \mathbf{R}_{A}}$ where the nucleus $A$ is in its original position.
The second term only lowers the variance, while the third term mainly fixes the bias when sampling by $\psi_T$ instead of $\psi_0$.
To summarize, both its systematic error and variance are reduced with better $\psi_T$ and better $\widetilde{\boldsymbol{\psi}}$, which is a huge improvement compared with the bare estimator.

Also, an approximation for $\widetilde{\boldsymbol{\psi}}_{A}$ is applied
\begin{equation}
    \widetilde{\boldsymbol{\psi}}_{A,\text{min}}(\mathbf{x})=\mathbf{Q}_{A}\psi_T
\end{equation}
where $\mathbf{Q}_{A}$ is a vector with $x,y,z$ components:
\begin{equation}
    \mathbf{Q}_{A}=Z_{A}\sum_{i=1}^{n_{\text{elec}}}\frac{\mathbf{r}_{i}-\mathbf{R}_A}{\left|\mathbf{r}_{i}-\mathbf{R}_A\right|}
\end{equation}
The approximation holds only when the electron approaches the nucleus $A$. Though it's a rough approximation, it's efficient and useful.

If the third term of equation \ref{general-O-estimator} can be ignored, we can further simplify the calculation by avoiding computing local energy $E_L$, which requires second order derivatives. For the sake of simplicity, we only write down the force on the nucleus of atom $A$ on $x$ axis:
\begin{equation}
    \begin{aligned}
    \widetilde{F}_{Ax,\text{AC-ZV}}
    &=F_{Ax}+\frac{\left(H-E_{L}\right)\widetilde{\psi}_{\text{min}}}{\psi_{T}}\\
    &=F_{Ax,\text{aa}}-\nabla Q_{Ax}\cdot\nabla\psi_{T}/\psi_{T}
    \end{aligned}
\end{equation}
This is what we call the AC-ZV estimator, because this estimator mainly lowers the variance. We call the full estimator from equation \ref{general-O-estimator} the AC-ZVZB estimator, because both variance and bias are lowered. And we call the third term the AC-Pulay term.
It's useful to treat the AC-ZV and AC-ZVZB estimators separately, because the AC-ZV estimator is a lot faster, and the bare estimator already has a bias of order $O[\delta\psi]$ and thus should give a relatively good result in a shorter time.

\subsubsection{Space Warp Coordinate Transformation estimator}

Another estimator is based on space-warp coordinate transformation (SWCT).
This method was initially introduced by Umrigar in 1989\cite{umrigar_two_1989}.
Filippi and Umrigar extended the idea in 2000\cite{filippi_correlated_2000},
and then Sorella and Capriotti implemented the method with adjoint algorithmic differentiation in 2010\cite{sorella_algorithmic_2010}.

The main idea is to displace electrons via a coordinate transformation according to the displacement of nuclei. If the electron and the nucleus are close enough, the electron will almost entirely follow the nucleus. To write it down
\begin{equation}
    \bar{\mathbf{r}}_i = \mathbf{r}_i + \Delta \mathbf{R}_{A} \omega_{A}(\mathbf{r}_i)
\end{equation}
\begin{equation}
    \omega_{A}(\mathbf{r})=\frac{f(|\mathbf{r}-\mathbf{R}_{A}|)}{\sum_{j=1}^{n_{\text{atom}}}f(|\mathbf{r}-\mathbf{R}_{j}|)}
\end{equation}
where $f$ is a fast-decaying function. Here we make the same decision as Filippi and Umrigar\cite{filippi_correlated_2000} and choose $f(r)=r^{-4}$.

Then the regularized force estimator is
\begin{equation}
    \widetilde{\mathbf{F}}_{A,\text{SWCT}}
    =-\frac{\mathrm{d}}{\mathrm{d} \mathbf{R}_{A}} E_{L}+
    2(E_v-E_L)\frac{\mathrm{d}}{\mathrm{d} \mathbf{R}_{A}} \log \left(|J^{1 / 2} \psi_T|\right)
\end{equation}
where $J$ is the Jacobian of the coordinate transformation.
We call the first term the Hellmann-Feynman term (SWCT-HFM) and the second SWCT-Pulay term.
And the HFM term can be expanded as
\begin{equation}
    \frac{\mathrm{d}}{\mathrm{d} \mathbf{R}_{A}} E_{L}=\frac{\partial}{\partial \mathbf{R}_{A}} E_{L}+\sum_{i=1}^{n_{\text{elec}}} \omega_{A}\left(\mathbf{r}_{i}\right) \frac{\partial}{\partial \mathbf{r}_{i}} E_{L}
\end{equation}
We will call the first part HFM-Bare term as it just considers local energy's derivative over nuclear position, and call the second part HFM-Warp term as it includes the effect of space warp.

If we omit the space warp effect and take $\omega_A=0$ and $J=1$, then we obtain the No-SWCT estimator, which was proposed by Sorella and Capriotti\cite{sorella_algorithmic_2010}.
\begin{equation}
    \widetilde{\mathbf{F}}_{A,\text{No-SWCT}}
    =-\frac{\partial}{\partial \mathbf{R}_{A}} E_{L}+
    2(E_v-E_L)\frac{\partial}{\partial \mathbf{R}_{A}} \log \left(|\psi_T|\right)
\end{equation}

\subsection{Implementation}
We implement the estimators above based on the JAX version of the FermiNet package \cite{pfau_ab_2020,spencer_better_2020,deepmind_ferminet_2022}.

The main loop for force calculation is straightforward and listed in Algorithm \ref{alg:main-loop}. Optionally, before the main loop, one can apply several Markov Chain Monte Carlo (MCMC) steps (called MCMC burn-in) to optimize electron configuration and reduce correlations between runs. Also, if the steps to run are not enough for an accurate $E_v$, one can add more inference steps (just computing local energy) to make the array storing local energies at each step (\texttt{all\_el}) long enough.

\begin{algorithm}[H]
    \caption{Calculate Force With Local Energy}
    \label{alg:main-loop}
    \begin{algorithmic}[1]
        \Require Network parameters \{\texttt{params}\}
        \Require MCMC Configuration \{\texttt{x}\}
        \State Initialize \texttt{hfm\_terms} (Hellmann-Feynman terms at each step), \texttt{el\_terms} (terms containing $E_L$), \texttt{ev\_term\_coeffs} (coefficient of $E_v$ term), \texttt{all\_el} (local energy) as arrays filled with zeros
        \State Initialize estimator with network function and system configuration
        \For{\texttt{i} $\gets 1,$ steps}
        \State \texttt{e\_l} $\gets$ local\_energy(\texttt{params}, \texttt{x})
        \State \texttt{hfm\_terms[i]}, \texttt{el\_terms[i]}, \texttt{ev\_term\_coeffs[i]} $\gets$ estimator(\texttt{params}, \texttt{x}, \texttt{e\_l})
        \State \texttt{all\_el[i]} $\gets$ \texttt{e\_l}
        \State \texttt{x} $\gets$ MCMC configuration after \texttt{n} steps
        \EndFor
        \State \texttt{forces} $\gets$ \texttt{hfm\_terms} + \texttt{el\_terms} + \texttt{ev\_term\_coeffs} $\times$ mean(\texttt{all\_el})
    \end{algorithmic}
\end{algorithm}

When implementing estimators, we express the equations above in Python JAX code \cite{jax2018github} for auto differentiation (AD) and calculating the derivative in all directions in one go. A tricky part when implementing the SWCT estimator is that we have to modify the local energy function to expose atom coordinates for differentiation and modify the potential energy part to get a well-defined gradient.

Also, it should be noted that all estimators in this work are sampled with plain $\psi_T^2$ instead of the reweighting scheme\cite{attaccalite_stable_2008}, which is included in the state-of-the-art traditional VMC calculations\cite{nakano_turborvb_2020,mazzola_unexpectedly_2014,mazzola_phase_2018,zen_abinito_2015}. That's because implementing the reweighting algorithm in FermiNet is not straightforward since FermiNet uses a linear combination of multiple determinants. Therefore, implementing and testing the reweighting algorithm in neural networks is left for future work.

Besides, when evaluating the SWCT estimator, some MCMC steps can give bad points, where the forces are ridiculously large.
To get a meaningful result, we apply the interquartile range (IQR) method to remove the outliers in each direction of the force result (\texttt{forces}) of all estimators.
The $\mathrm{IQR}$ is the difference between the 25th percentile ($Q_1$) and the 75th percentile ($Q_3$) in a dataset.
All data greater than $Q_3 + 3  \mathrm{IQR}$ or less than $Q_1 - 3 \mathrm{IQR}$ is clipped to the boundaries.
$3 \mathrm{IQR}$ is chosen instead of the popular $1.5 \mathrm{IQR}$ because we want to have fewer data clipped.
We also tested a different clipping scheme, where the top 3\% and the bottom 3\% are clipped off. 
The results using different clipping schemes are almost identical, and the data reported in this work are with the IQR method.

\subsection{Computational settings}

There are various settings in the training and calculation process. In the training part, we use the default config recommended by the FermiNet package\cite{pfau_ab_2020,spencer_better_2020}. The detailed setup is listed in Table. \ref{tab:settings}.
The accuracy of wavefunction with respect to network size has been discussed in other works (e.g. Refs. \onlinecite{pfau_ab_2020,spencer_better_2020,ren_towards_2022}), in this study the performance of force estimators and the dependence on the quality of wavefunction are discussed by varying the number of training iterations.

\onecolumngrid
\begin{center}
\begin{table}[h]
\setlength{\tabcolsep}{6pt}
\caption{Computational settings}
\label{tab:settings}
\begin{tabular}{@{}clp{5.5cm}@{}}
\toprule
Source                     & Name                          & Value \\
\colrule
\multirow{10}{*}{FermiNet}
& Framework                     & JAX\cite{jax2018github} \\
& Hardware                      & 16GB(32GB) V100 GPU card(s) \\
& Batch size                    & 4096  \\
& Hidden units per one-electron layer  & 256 \\
& Hidden units per two-electron layer  & 32 \\
& Number of layers              & 4 \\
& Number of determinants        & 16 \\
& Train iterations              & Default $10^5$ for $\mathrm{H}_2$, and $2\times 10^5$ for $\mathrm{Li}_2$ and $\mathrm{N}_2$. Different quality of network is achieved by training with different iterations. \\
& Optimizer                     & KFAC-JAX\cite{deepmind_kfac-jax_2022} \\
& Other settings                & Default\cite{pfau_ab_2020,deepmind_ferminet_2022} \\
\colrule
\multirow{3}{*}{This work}
& MCMC burn-in steps            & 100   \\
& MCMC steps between iterations & 50    \\
& Extra energy inference steps  & 0     \\
\botrule 
\end{tabular}
\end{table}
\end{center}
\twocolumngrid

\section{Results and discussion}
\subsection{Interatomic force from well converged neural network}
\label{subsec:force-result}

\begin{figure*}[htbp]
    \includegraphics[width=0.8\textwidth]{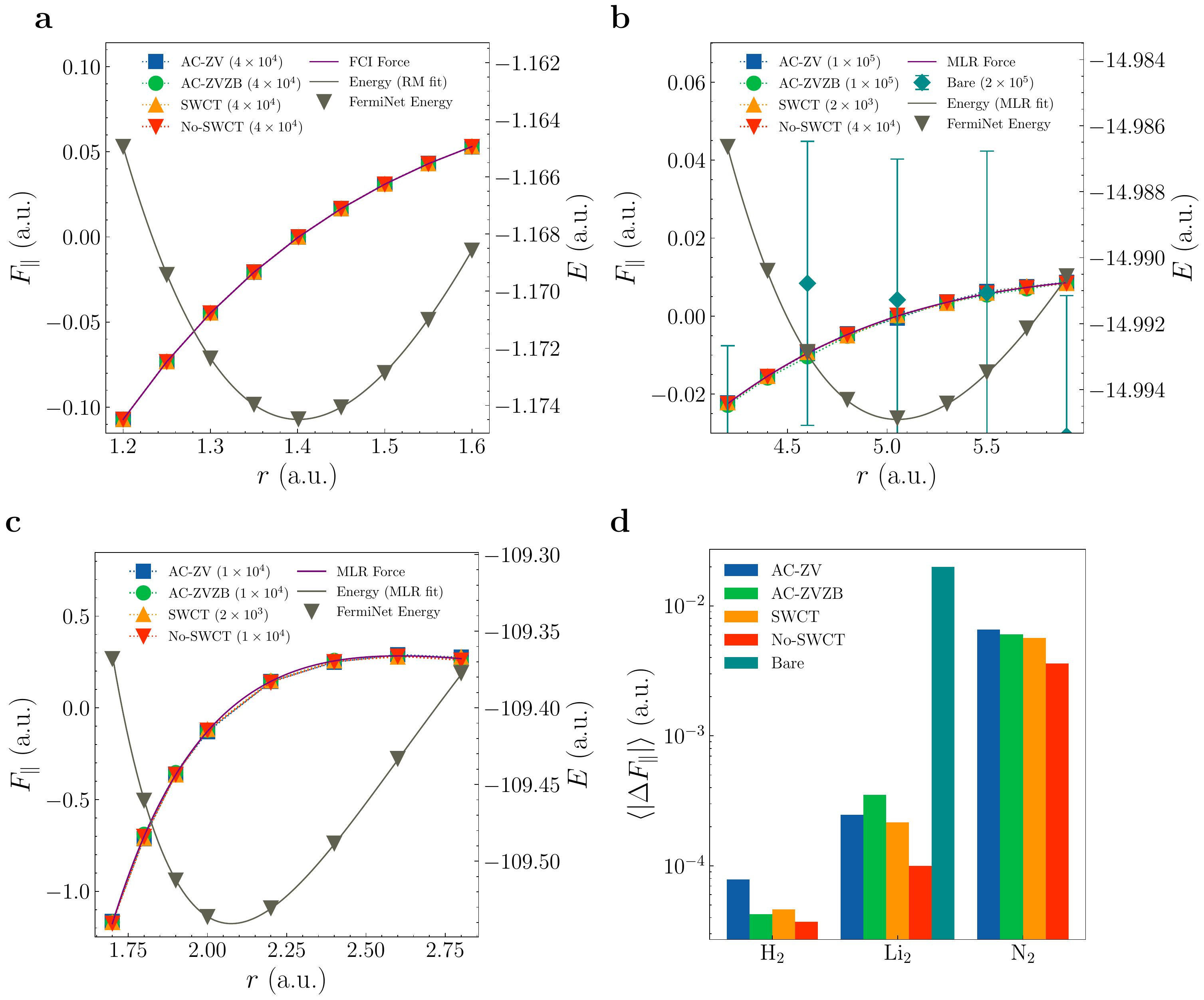}
    \caption{
        \textbf{(a)} to \textbf{(c)} Force ($F_\parallel$) curves of (a) $\mathrm{H}_2$, (b) $\mathrm{Li}_2$, and (c) $\mathrm{N}_2$ molecules with well-converged neural networks.
        The $\parallel$ subscript is used to stress that it's the parallel component with the molecule orientation.
        In each panel are the energy curve (grey lines and down triangles symbols) and the force curves (other lines) which are in excellent agreement with each other.
        The forces estimators include the AC-ZV (blue square), the AC-ZVZB (green circle), the SWCT (orange up triangle), the No-SWCT (red down triangle), and the bare (dark cyan diamond) estimators.
        The brackets indicate the number of force calculation steps.
        The purple lines are reference data for comparison including FCI calculations of $\mathrm{H}_2$ with the cc-pV6Z basis set, and the derivative of Morse/Long-range potentials of $\mathrm{Li}_2$ and $\mathrm{N}_2$\cite{le_roy_accurate_2009,le_roy_accurate_2006}. 
        \textbf{(d)} The average differences between results from different estimators and the reference data over different bond lengths ($\langle |\Delta F_\parallel| \rangle$). 
    }
    \label{fig:well-trained-result}
\end{figure*}

The force component parallel to the direction of molecular orientation of $\mathrm{H}_2$, $\mathrm{Li}_2$, and $\mathrm{N}_2$ molecules at different bond lengths are shown in Fig. \ref{fig:well-trained-result}.
The results are based on well-trained neural networks, and the number of steps are chosen so that results from different estimators (except for the bare estimator) have similar statistical errors.
The main message here is that with the regularized estimators the force curves can be accurately calculated.
The forces from different regularized estimators are all in excellent agreement with each other and have small statistical errors.
In contrast, the bare estimator has much larger systematic and statistical errors (Fig. 1b), which was originally recognized in traditional VMC calculation, and thus the bare estimator is excluded from further discussions.
To further evaluate whether there is a systematic bias in our force estimates, we also plot data from other calculations for comparison.
In the case of $\mathrm{Li}_2$ and $\mathrm{N}_2$ molecule, the reference data is taken by the derivative of an accurate Morse/Long-range potential \cite{le_roy_accurate_2009,le_roy_accurate_2006}.
For $\mathrm{H}_2$, such a reliable analytical bonding curve is not available so we calculated another force curve using the full configuration interaction (FCI) method on the cc-pV6Z basis set.

To visualize the order of magnitude of the minor differences between our calculations and the reference, we plot the average difference between our FermiNet-VMC force results and the selected reference data in Fig. \ref{fig:well-trained-result}d.
We find the differences are very small.
For tiny systems like $\mathrm{H}_2$ and $\mathrm{Li}_2$, the number of calculation steps is large enough so that the differences are no more than $10^{-3}$ a.u.
Although the taken references might also have bias, these comparisons convincingly show that the error of the FermiNet-VMC force is less than $10^{-3}$ a.u. in these two systems.
We note that in previous studies with a traditional wavefunction ansatz, the VMC force error on $\mathrm{H}_2$ and $\mathrm{Li}_2$ can be as large as $10^{-1}$ a.u.\cite{assaraf_zero-variance_2003}, which is two orders of magnitude larger.
This is mainly because of the simple wavefunctions used, and our calculations further confirm that these estimators' zero-variance and zero-bias properties can benefit from the expressiveness of neural network wavefunction ansatz.
For the larger $\mathrm{N}_2$ molecule, the neural network training is much more time consuming, and our force results are computed with an insufficient number of training steps,
thus the errors are at the order of sub $10^{-2}$ a.u.
The results can be improved when the training of the neural network is improved, as we will show below.

A more detailed comparison of different estimators and more discussions about the cost of different force estimators will also be presented below.

\subsection{Dependence on the quality of neural network}
\label{subsec:network-dependence}
\begin{figure*}[htbp]
    \includegraphics[width=0.95\textwidth]{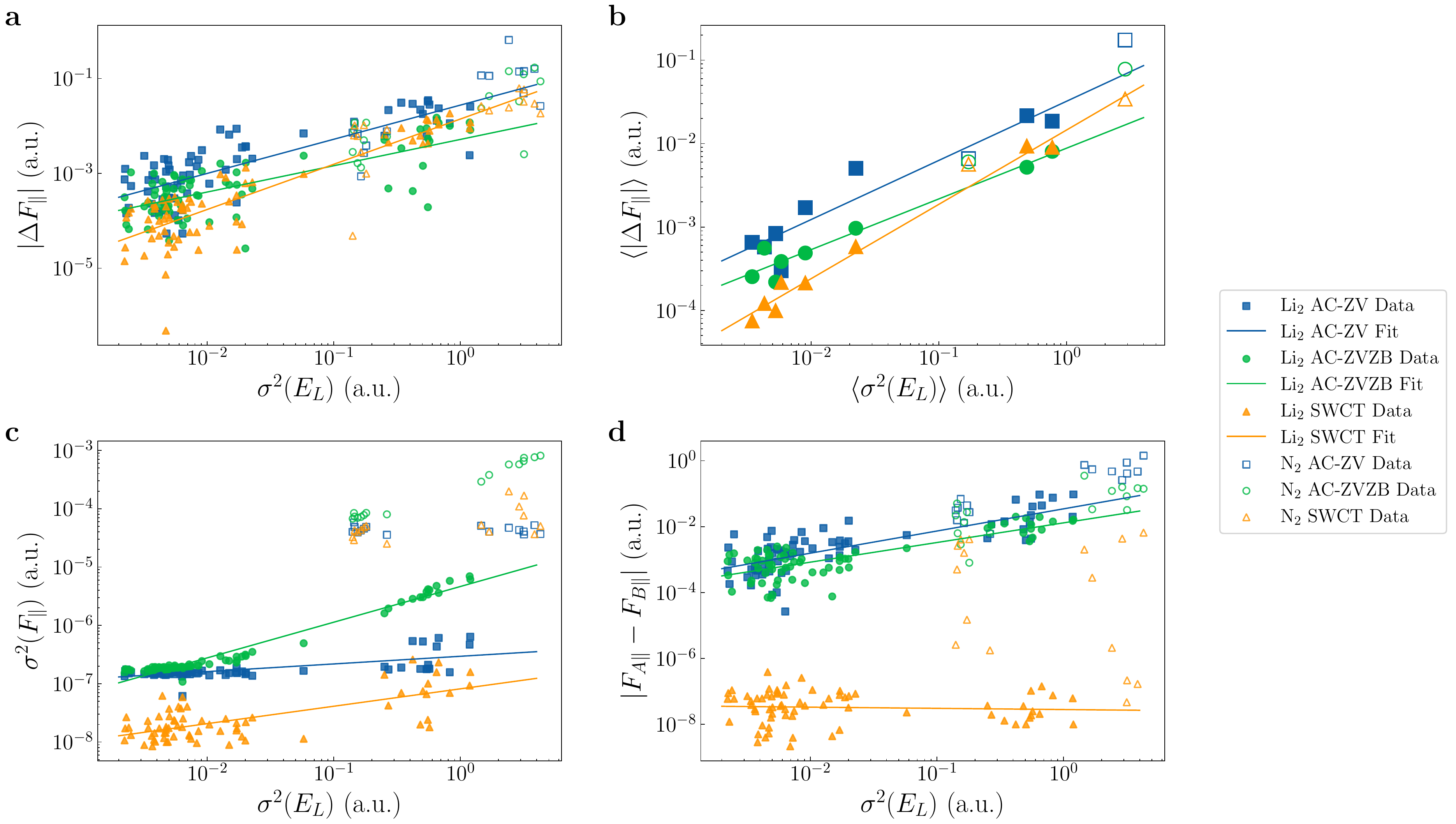}
    \caption{
        \textbf{(a)} The relation between the absolute value of the force error ($|\Delta F_\parallel|$) and the energy variance ($\sigma^2 (E_L)$) with different $\rm{Li}_2/\rm{N}_2$ bond lengths.
        The force error is defined as the difference between our calculation and the MLR force reported in Fig. 1.
        \textbf{(b)} Similar to (a) but the errors and the variances are averaged over different bond lengths for checkpoints with the same number of training steps ($\langle |\Delta F_\parallel| \rangle$, $\langle \sigma^2 (E_L) \rangle$).
        \textbf{(c)} The relation between force variance ($\sigma^2 (F_\parallel)$) and energy variance ($\sigma^2 (E_L)$).
        \textbf{(d)} The relation between force difference between nucleus $A$ and $B$ ($|F_{A\parallel}-F_{B\parallel}|$) and energy variance ($\sigma^2 (E_L)$).
        All AC-ZV and AC-ZVZB results are calculated with $10^4$ steps, and SWCT results are calculated with $2\times10^3$ steps.
        The energy variance is obtained by using equation \ref{eq:std-mean}, and the error is obtained by comparing with the derivative of Morse/Long-range results\cite{le_roy_accurate_2009,le_roy_accurate_2006}.
        And the lines are obtained by fitting $\rm{Li}_2$ data points with least squares.
    }
    \label{fig:something-vs-energy-variance}
\end{figure*}

We have discussed the performance of estimators when the neural network is well-trained in the above section.
To examine whether there is a relation between the performance of the force estimators and the quality of the neural network, i.e. the accuracy of the trial wavefunction, we have calculated forces with various neural network checkpoints at different training steps for the $\mathrm{Li}_2$ and $\mathrm{N}_2$ molecules and the results are shown in Fig. \ref{fig:something-vs-energy-variance}.
The energy variance is the equivalent variance of local energies ($\sigma^2(E_L)$), 
converted from the standard error $\sigma_{\mathrm{block}}(E_L)$ given by the blocking algorithm\cite{hesse_analysis_2021} by using the standard deviation formula for the mean:
\begin{equation}\label{eq:std-mean}
    \sigma^2(E_L) = \frac{1}{2\tau_{\text{int}}} \cdot n_{\mathrm{steps}} \cdot n_{\mathrm{batch}} \cdot \sigma_{\mathrm{block}}^2 (E_L)
\end{equation}
where $\tau_{\text{int}}$ is the integrated autocorrelation time\cite{wolff_monte_2004} for $E_L$, and $n_{\mathrm{steps}}$ is the number of calculation steps, and $n_{\mathrm{batch}}$ is the batch size, i.e. the number of local energies calculated in one step.

In Fig. \ref{fig:something-vs-energy-variance}a we plot the error on force as a function of the energy variance of FermiNet-VMC, which illustrates a correlation between the quality of the trial wavefunction and the accuracy of force estimation. In Fig. \ref{fig:something-vs-energy-variance}b the data are averaged over different bond lengths for checkpoints with the same number of training steps, which shows the same trend but much more evidently.
Among the three estimators, we note that the force error of the SWCT estimator reduces the fastest when the network becomes better trained, and achieves the lowest error compared with those of AC-ZV and AC-ZVZB estimators when the network is well-trained.
However, the average error of the SWCT estimator can be larger when the network is undertrained.
In addition, we note that the AC-ZVZB estimator generally gives a better result than the AC-ZV estimator.

Apart from the error, the variance of the force also can benefit from the quality of the trial wavefunction, as shown in Fig. \ref{fig:something-vs-energy-variance}c where we plot the relation between the variance of force and the energy variance.
It is easy to notice that the SWCT estimator always produces the lowest force variance, which is consistent with what we found in the previous subsection.

Another noteworthy point is that due to the statistical error, the final force result is not the same on the two nuclei, and the difference can be quite large for AC-type estimators.
In Fig. \ref{fig:something-vs-energy-variance}d we plot the force difference between nucleus $A$ and $B$ as a function of the energy variance.
For the AC type estimators, the force differences between two nuclei are notably reduced when the quality of the network improves.
When the trial wavefunction is good enough, the AC type estimators give a force difference of roughly one-tenth of the final force results.
On the contrary, SWCT gives negligible difference between forces on the two nuclei, which is under $10^{-7}$ a.u. for all the trained neural networks of $\mathrm{Li_2}$.

We also plot the results for $\mathrm{N_2}$ molecule in Fig. \ref{fig:something-vs-energy-variance}, which show that the trends of the $\mathrm{Li}_2$ molecule in panels (a), (b) and (d) are still applicable to the $\mathrm{N_2}$ molecule.
This suggests that one can roughly estimate the force bias from the local energy variance, which should be useful in studying larger molecules where accurate benchmark data of interatomic force are not available. 
Specifically, in atomic units, the force bias is about one percent of the local energy variance.
For geometry optimization, molecular dynamics, and force field training, it is often required to have interatomic force within an error of $10^{-3}$ a.u., $10^{-2}$ a.u., and $10^-2$ a.u., respectively.
Therefore, when such tasks are aimed, a simple criterion for neural network training is to reduce the variance of local energy within $10^{-1}$ a.u., $10^0$ a.u., and $10^0$ a.u., respectively.
Regarding the force variance (Fig. 2c), although the variance of $\mathrm{N_2}$ also decreases as the network is better trained, the data are not in line with the fitting curve of $\mathrm{Li_2}$, therefore the statistical error of force should be estimated for each specific calculation.

\subsection{Orthogonal components}
\label{subsec:orthogonal}

\begin{figure}[tbp]
    \includegraphics[width=0.8\columnwidth]{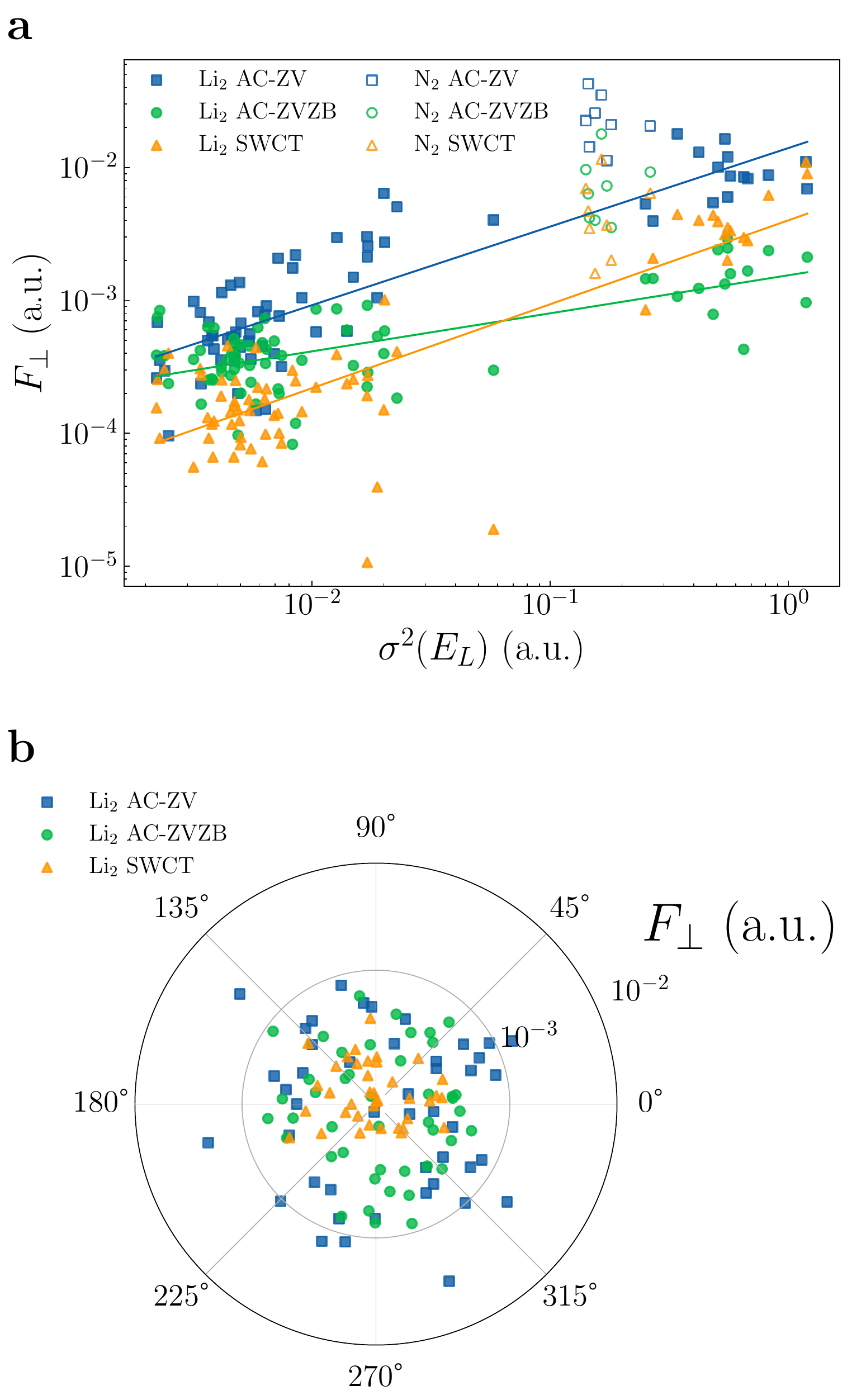}
    \caption{
        \textbf{(a)} The relation between the orthogonal force component ($F_\perp$) and the energy variance ($\sigma^2 (E)$) of all $\rm{Li}_2/\rm{N}_2$ bond lengths.
        \textbf{(b)} The direction and the magnitude of the orthogonal force component of different $\mathrm{Li}_2$ bond lengths at various neural network checkpoints.
        }
    \label{fig:orthogonal}
\end{figure}

In the above subsections, we have discussed the results of the parallel force component.
The orthogonal component of the interatomic force is supposed to be exactly zero for diatomic molecules, 
but due to statistical errors they can also be non-zero values.
In this subsection, we show that the orthogonal component gives similar results as the parallel component.
Like subsection \ref{subsec:network-dependence}, we plot the magnitude of the orthogonal force component as a function of the energy variance in Fig. \ref{fig:orthogonal}a.
By comparing Fig. \ref{fig:orthogonal}a with Fig. \ref{fig:something-vs-energy-variance}b, we can find that the error of orthogonal and parallel components are of the same order of magnitude under the same network quality.
Similarly, the error of the orthogonal component can be reduced by improving the quality of the network.

In Fig. \ref{fig:orthogonal}b, we plot the distribution of the direction and the magnitude of the orthogonal component, which shows that the direction is randomly distributed.
This confirms that the error of the orthogonal force component is mainly composed of statistical errors.

\subsection{Contribution of different terms}
\label{subsec:terms}
\begin{figure}[tbp]
    \includegraphics[width=0.8\columnwidth]{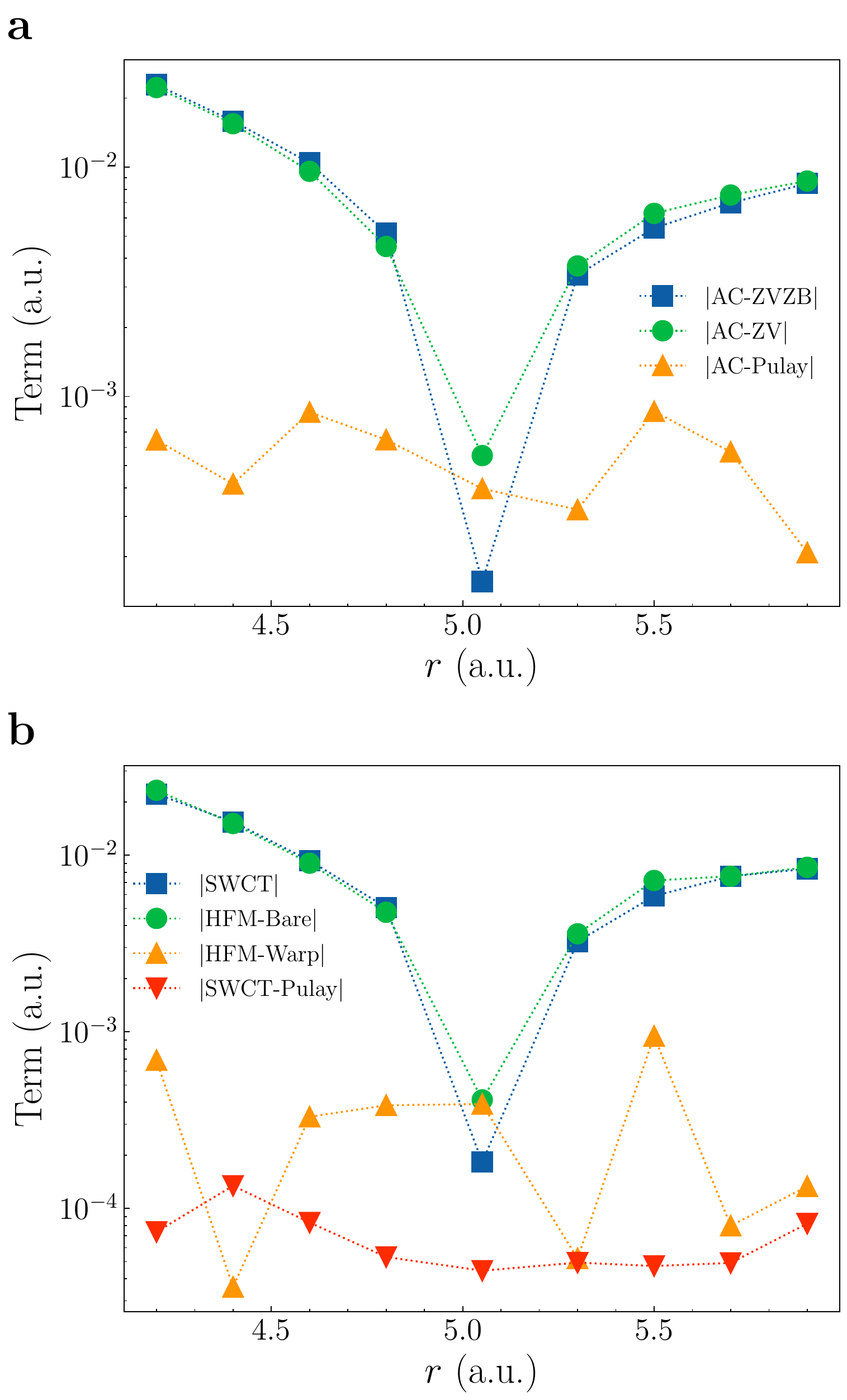}
    \caption{
        \textbf{(a)} Absolute value of the AC-ZV, AC-ZVZB estimation, and its Pulay term at each bond length of $\mathrm{Li}_2$ molecule, calculated with $10^5$ steps.
        \textbf{(b)} Absolute value of the SWCT estimation, the HFM-Bare term, the HFM-Warp term, and the SWCT-Pulay term at each bond length of $\mathrm{Li}_2$ molecule, calculated with $2\times 10^3$ steps.
        }
    \label{fig:terms-contribution}
\end{figure}

Apart from the total force results, we can further examine the contribution of different terms of the AC-ZVZB and SWCT estimators, as shown in Fig. \ref{fig:terms-contribution} with the example of $\mathrm{Li}_2$. 
In Fig. \ref{fig:terms-contribution}a, we plot the AC-ZVZB, AC-ZV, and AC-Pulay force curves, where the last term captures exactly the difference between the first two estimators.
We find the AC-Pulay term, which was designed to reduce the systematic error of force calculation, is on the order of $10^{-4}$ to $10^{-3}$ a.u.
At the same time, we can see from Fig. 1 that both the statistical error and the difference between different estimators are on the same order of magnitude.
Thus, the Pulay term becomes non-negligible when using the Assaraf-Caffarel estimators, especially when the geometry is near equilibrium.
When the atoms are near the equilibrium positions, the dominating term (AC-ZV) is close to zero, thus the percentage of the Pulay term is the highest.
When the atoms move away from the equilibrium, the percentage of the Pulay term decreases rapidly.

The contributions of different terms in the SWCT estimator are shown in Fig. \ref{fig:terms-contribution}b. 
Similar to the AC-ZVZB method, there is one dominant term (HFM-Bare) when the atoms are away from the equilibrium positions.
The HFM-Warp and the SWCT-Pulay terms are comparably smaller by one or two orders of magnitude.
The Pulay term is the smallest and is around $10^{-4}\:\mathrm{a.u.}$ at all bond lengths, meaning that with the SWCT method the Pulay term can be neglected.
The HFM-Warp term is slightly larger and should be considered when high precision is needed.


\subsection{Time per step}

\label{subsec:timing}
\begin{figure}[tbp]
    \includegraphics[width=0.8\columnwidth]{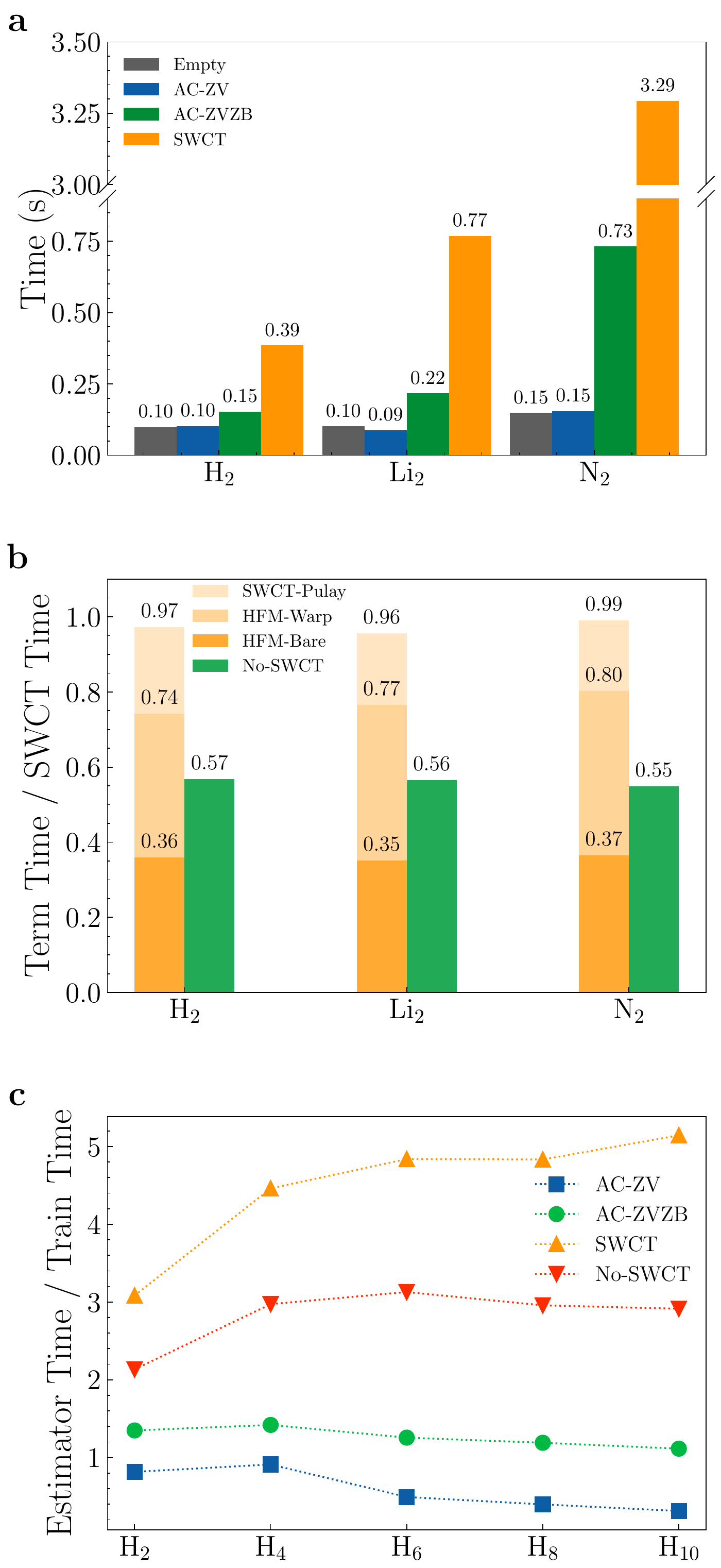}
    \caption{
        \textbf{(a)} Computational time for each step of AC-ZV, AC-ZVZB, SWCT estimators, and an empty loop.
        \textbf{(b)} Computational time of different terms of the SWCT estimator, divided by the time of a full SWCT step, excluding the time of an empty loop, i.e. $(t_\mathrm{term} - t_\mathrm{empty})/(t_\mathrm{SWCT}-t_\mathrm{empty})$.
        \textbf{(c)} The relative cost (ratio between force estimator calculation time and training time per step) of AC-ZV, AC-ZVZB, SWCT, and No-SWCT estimators for $\mathrm{H}_{2}$ to $\mathrm{H}_{10}$.
        Different hardware settings are used for different molecules, as listed in Table \ref{tab:time-hardware}
        }
    \label{fig:time}
\end{figure}

A comparison of time taken by a single step of all three estimators is shown in Fig. \ref{fig:time}a.
An empty loop means that neither force nor local energy is calculated, and the simulation contains 50 MCMC steps between two force calculation steps.
For a single step, the AC-ZV estimator runs roughly as fast as the empty loop, which means that it actually requires a negligible amount of calculations compared with MCMC steps.
Then follows the AC-ZVZB estimator, and the SWCT estimator is the slowest.
This is the direct consequence of the order of derivative the estimators require: the AC-ZV estimator requires the first order, AC-ZVZB requires the second, and SWCT requires the third.
However, considering that much fewer steps are required for the SWCT estimator to perform similarly to the AC-ZV or AC-ZVZB estimator, for the systems tested in this work the SWCT method is actually more efficient.

Since the SWCT estimator takes so much time, it is worth looking into different terms' time costs of the estimator and finding out  which part takes the most.
The result is shown in Fig. \ref{fig:time}b.
The bars of HFM-Bare, HFM-Warp term, and Pulay terms add up to approximately 100\%, with some statistical errors.
The SWCT-HFM term takes the majority of the time, with the bare term and the warp term each taking approximately half of it.
And the Pulay term takes around a quarter to a fifth of the total cost.
As the system grows, the SWCT-HFM term takes a larger portion of the total cost, which is understandable since it requires third-order derivatives while the Pulay term only requires the second order.
We also take a look at the No-SWCT estimator, and we find that it takes around half of the time of the SWCT estimator.

Another trend we can find is that the ratio between the speed of estimators is not the same for different molecules.
The AC-ZVZB and the SWCT estimator slow down much faster than the AC-ZV estimator as the system grows larger and more electrons are involved.
That's a consequence of the different complexities of the estimators.
As Pfau et al. pointed out\cite{pfau_ab_2020}, when using AD, evaluating the determinants scales as $O(n_{\text{elec}}^3)$, and evaluating the gradient of a function has the same asymptotic complexity as the function, and evaluating the local energy contributes an additional factor of $n_{\text{elec}}$.
Thus the AC-ZV estimator scales as $O(n_{\text{elec}}^3)$ just like evaluating the determinants, and the AC-ZVZB and the SWCT estimator scales as $O(n_{\text{elec}}^4)$ just like evaluating the local energy.
We further test the estimators on linear hydrogen systems including $\mathrm{H}_2$, $\mathrm{H}_4$, $\mathrm{H}_6$, $\mathrm{H}_8$, and $\mathrm{H}_{10}$ molecules, and the results are plotted in Fig. \ref{fig:time}c.
As shown in the figure, the relative costs of AC-ZVZB, SWCT and No-SWCT estimators remain steady from $\mathrm{H}_4$ to $\mathrm{H}_{10}$, similar to what has been shown by Sorella et al. in traditional VMC code \cite{sorella_algorithmic_2010}. And the relative cost of the AC-ZV estimator goes down when the number of hydrogen atoms grows.
This feature is very useful for the application of force modules within FermiNet-VMC in the future.

\subsection{Memory consumption}

\begin{figure}[tbp]
    \includegraphics[width=0.8\columnwidth]{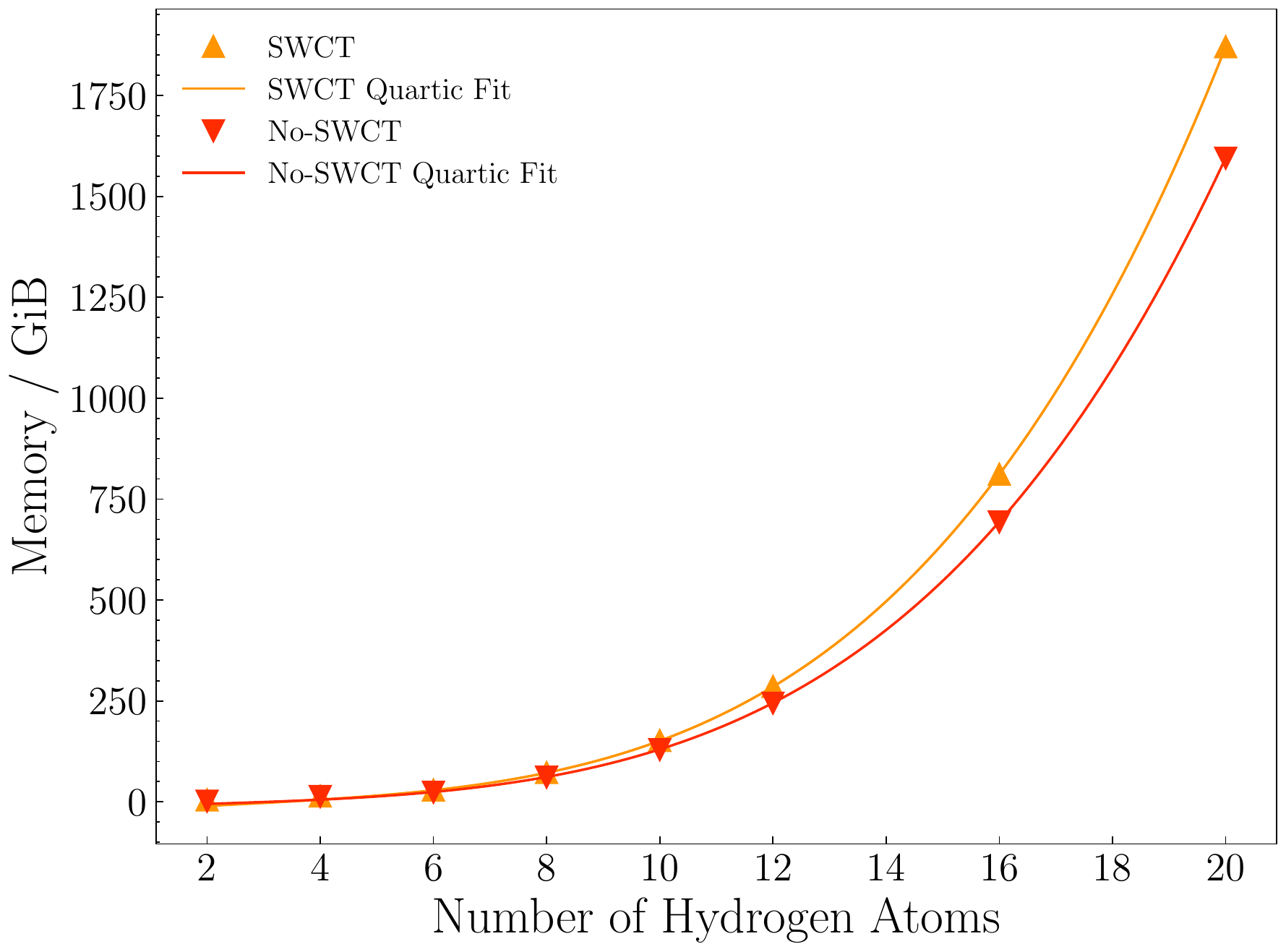}
    \caption{
        GPU memory consumed by SWCT and No-SWCT estimators for $\mathrm{H}_n$ systems. 
        }
    \label{fig:swct-memory}
\end{figure}

AC-ZV and AC-ZVZB estimators require similar memory as FermiNet itself, but that's not the case for the SWCT and No-SWCT estimators.
Since the gradient of local energy $E_L$ is needed, and we are doing this through reverse mode of AD, the memory required by SWCT and No-SWCT is notably higher.
We have to use more GPU cards with more memory to make it possible to evaluate all 4096 configurations in one go. 
In Fig. \ref{fig:swct-memory} we plot an estimate of the GPU memory required for $\mathrm{H}_2$ to $\mathrm{H}_{20}$ systems, where $n_{\text{elec}}=n_{\text{atom}}=n$.
For $\mathrm{H}_6$ to $\mathrm{H}_{20}$, the memory required is estimated from the out of memory error message.
For $\mathrm{H}_2$ and $\mathrm{H}_4$, the memory cost is estimated from hardware statistics, which might be slightly overestimated.
Within the studied regime, the memory cost has an $O(n^4)$ scaling, which increases quickly the memory requirements for large systems.
If such resources are not available, one can split the configurations into several chunks and evaluate one chunk at a time, or even one configuration at a time, to overcome memory issues, at the sacrifice of efficiency. 

\section{Conclusions}

We have implemented the AC-ZV and AC-ZVZB estimators suggested by Assaraf et al.\cite{assaraf_zero-variance_2003}, and the SWCT estimator developed by Umrigar et al.\cite{umrigar_two_1989,filippi_correlated_2000,sorella_algorithmic_2010} based on FermiNet-VMC, and calculated forces of $\mathrm{H_{2}}$, $\mathrm{Li_{2}}$, and $\mathrm{N_{2}}$ along their potential energy curves.
We compared the estimators on both well-trained and undertrained network wavefunction and found that the quality of neural network benefits the accuracy of force significantly, which indicates the promising future of the application of neural network wavefunction methods in force field developments, structure optimization, and molecular dynamics.
Among the different estimators, the SWCT estimator is the most accurate and achieves the best accuracy-cost balance when the system is not too large.
For calculations on large systems with low accuracy requirements, the AC-ZV estimator is a more efficient choice.

\section*{Data availability}
The data that support the findings of this study are available from the corresponding author upon reasonable request.

\begin{acknowledgments}
This work was supported by the National Key R\&D Program of China under Grant No. 2021YFA1400500, the National Natural Science Foundation of China under Grant No. 11974024 and No. 92165101, and the Strategic Priority Research Program of Chinese Academy of Sciences under Grant No. XDB33000000. The authors thank the High Performance Computing Platform of Peking University for computational resources. We also thank Hang Li and ByteDance AI-Lab for support and helpful discussions.
\end{acknowledgments}

\bibliography{ref.bib}

\end{document}